\documentclass[twocolumn,aps,prl,showpacs,preprintnumbers,amsmath,amssymb]{revtex4}

\usepackage{graphicx}
\usepackage{dcolumn}
\usepackage{bm}
\newcommand{\Ang}{\,\mathrm{\AA}}

\newcommand{\ev}{\,\mathrm{eV}}

\begin{document}
%{\parbox[b]{1in}{\hbox{\tt INT-PUB-09-012}}}
\title{Correlation between Charge Inhomogeneities and Structure in Graphene and Other Electronic Crystalline Membranes}
\author{Doron Gazit}
\email{doron.gazit@mail.huji.ac.il}
\affiliation{Institute for Nuclear Theory, University of Washington, Box 351550, Seattle, WA 98195, USA}

\date{\today}

\begin{abstract}
Only one atom thick and not inclined to lattice defects, graphene represents the ultimate crystalline membrane. However, its structure reveals unique features not found in other crystalline membranes, in particular the existence of ripples with wavelength of $100-300\Ang$. Here, I trace the origin of this difference to the free electrons in the membrane. The deformation energy of the lattice creates a coupling between charge fluctuations and the structure, resulting in ripples on the membrane, correlated with charge inhomogeneities. In graphene this mechanism reproduces the experimental result for both charge puddles and ripples. %The same mechanism also gives a simple formula for the elctron-phonon coupling in superconductors, depending only on the elastic constants and the deformation energy, and hints to the correlation found between charge inhomogeneities and lattice deformations in some layered superconductors.
\end{abstract}


\maketitle
Physical membranes are objects in which one of the dimensions is small compared to the other two, giving them an effective two-dimensional (2D) character. An intriguing class of physical membranes is that of the crystalline membranes, that have a solid structure, usually of a triangular or hexagonal 2D lattice \cite{Membranes_Book}. 

Beautiful examples of such systems, that spread in
magnitude and scale, exist in our world. 
In biology, a famous example is the cytoskeleton of red blood-cells, whose structure is vital for the operation and stability of the cell, that forms a triangular crystalline lattice built of spectrin proteins. In soft condensed matter, one can create crystalline lattices by polymerizing liquid interfaces \cite{Membranes_Book}.  
Recently, an ingenious experimental method, using mechanical cleavage, has
conquered the final limit, isolating graphene -- a single-layer of carbon atoms, organized in a hexagonal lattice \cite{2004SciNovoselov}. The same method has since been used to isolate individual crystal planes of other layered materials \cite{2005PNASNovoselov}.

With such a vast variety of crystalline membranes in nature, it is of no surprise that understanding the structural properties of these systems has attracted the attention of many physicists \cite{Membranes_Book,2001PhRBowick,1987JPhNelson,1988EPLDavid_1,1988EPLDavid_2, 1988PRLAronovitz, 1992PRLSCSA, 1993SciSchmidt, 1996JPhBowick,2009ArXivGazit}. The fact that graphene can be used to construct nanometer-sized electronic applications, has only enhanced the need of a profound understanding of its structure \cite{2007NatMatGeim}.
This interest intensifies, in view of the Mermin-Wagner theorem, which forbids the
existence of long range order in 2D systems due to diverging thermal
vibrations. This seeming contradiction between experiment and theory, which peaked with the discovery that graphene is stable even when it is free standing, i.e., without the support of a substrate \cite{2007NatMeyer}, is resolved by allowing out-of-plane fluctuations,
%\cite{Membranes_Book,2001PhRBowick,1987JPhNelson,1988EPLDavid_1,1988EPLDavid_2, 1988PRLAronovitz, 1992PRLSCSA, 1993SciSchmidt, 1996JPhBowick}.
that induce frustration between the large thermal vibrations in 2D and the competing gain in elastic
energy. This frustration stabilizes a globally flat phase at finite temperatures \cite{Membranes_Book}. This ``almost flat" phase is characterized by a scale invariant structure, at wavelengths much longer than a characteristic size $\lambda_T=\frac{2\pi \kappa}{\sqrt{K_0k_BT}}$, determined by the bending energy  $\kappa$, and the 2D-Young modulus $K_0$ ($T$ is the temperature) \cite{Membranes_Book,2001PhRBowick,1987JPhNelson,1988EPLDavid_1,1988EPLDavid_2, 1988PRLAronovitz, 1992PRLSCSA, 1993SciSchmidt, 1996JPhBowick}. However, experimental studies of graphene have revealed rather different features. 

Meyer et al \cite{2007NatMeyer} have isolated a free-standing graphene, thus demonstrating its long-range-order and stability. Furthermore, they used transmission electron spectroscopy to study its structure. Their finding, which was since reproduced by other experimental groups, is that the graphene sheet exhibits spontaneous rippling, with amplitude of about $3-10 \Ang$, and wavelength estimated to be $\lambda=100-300 \Ang$ \cite{2007NatMeyer,2007ApplPhysLettBrar,2009PhysRevLettGeringer}. Lattice defects were not found, and thus cannot account for this non-vanishing curvature \cite{2007NatMeyer,2005NatZhang}. Considering the fact that in graphene $\lambda_T\approx12\Ang$, these ripples clearly violate the scale invariance that should govern at this scale. 

As graphene is the ultimate crystalline membrane, this difference has to be addressed theoretically. 
In this {\it Rapid Communication}, I suggest that the ripples in graphene are a signature of the fact that it is not a regular crystalline membrane, since it has an additional degree of freedom -- the free electrons that occupy its $\pi$ band -- thus it is a representative of a different class of materials -- electronic crystalline membranes, in which an interplay exists between the electronic and structural degrees of freedom \cite{2008EPLKim,2009PhysRevBGazit}. This interplay leads to the excitation of ripples in electronic crystalline membranes in general, and graphene in particular.

To reach this conclusion, I start by modeling the ``almost-flat'' phase of a membrane. In-plane
deformations are characterized by a two dimensional vector field $\vec{u}$, and out-of-plane deformations by a field $h$. When considering the equilibrium state of the electronic crystalline membrane, without allowing charge fluctuations in the conduction electrons, the mesoscopic structure of the membrane can be described by thermal fluctuations around this equilibrium using the elastic free energy:
\begin{eqnarray} \label{Eq:Elastic_energy}
F[u,h] =\frac{1}{2}\int d^2\vec{x} \left[\kappa (\nabla^2 h)^2 + 
2\mu  u_{ij}^2 +\lambda u_{ii}^2 \right],
\end{eqnarray}
where $u_{ij}$ is the strain tensor,
$u_{ij}=\frac{1}{2}\left(\partial_i u_j+\partial_j u_i\right)+\frac{1}{2}(\partial_i h)(\partial_j h)$ (summation over repeated indices is implied throughout the manuscript). The coefficients are the 2D elastic properties of the membrane. In graphene, experiments have verified this approximation, as no lattice defects were found even at strains $\sim 10\%$\cite{2007NatMeyer,2005NatZhang}, and the elastic constants were estimated: a bending energy $\kappa \approx 1.1 \ev$, bulk modulus $\lambda+\mu \approx 7.3 \ev  \Ang^{-2}$, and shear modulus $\mu \approx 5.7 \ev  \Ang^{-2}$ \cite{2009SolStaCommGuinea} (the resulting Young modulus is $K_0=\frac{4\mu(\mu+\lambda)}{2\mu+\lambda}\approx 13 \ev \Ang^{-2}$). $\mu$ and $\lambda$ were estimated from the sound velocities. These elastic properties originate in the $\sigma$ band, which is a consequence of the in-plane $sp^2$ hybridization, that forms a deep valence band, and partly by the $\pi$ band, which is perpendicular to the plane. This was also verified by Monte-Carlo simulations with realistic, though phenomenological, inter-atomic potentials (that cannot take into account charge fluctuations) \cite{2007NatMaFasolino,2009PhysRevLettZakharchenko,2009PhysRevBFasolino}. An additional verification that Eq.~(\ref{Eq:Elastic_energy}) indeed describes such simulations, is due to the fact that the structure of graphene predicted by them exactly fits the theory of crystalline membranes, quantitatively predicting the scale invariance and the anomalous exponents of the bending energy and elastic constants at long wavelengths \cite{2009PhysRevBFasolino}, however unable to reproduce the ripples. Thus, the ripples have a different origin. 

The free electrons, which in graphene occupy the half-filled $\pi$ band, differentiate an electronic crystalline membrane from regular crystalline membranes. Indeed, the free electrons couple to the structure through a deformation energy. The source of this deformation potential is the local change in the Fermi energy measured from the bottom of the valence band, proportional to the local change in area $\delta S$. 
The resulting deformation potential has the form $V_s=D\frac{\delta S}{a^2}=D{u_{ii}}$ ($a$ is the nearest neighbor distance), where $D$ is the Fermi energy of the 2D electron gas \cite{2002PhysRevBAndo}. In graphene this deformation energy was found to be the main source of deviations of the electrical transport properties from ballistic transport. The specific value of $D$ is a matter of debate. However, using charge carrier mobility measurements of electron doped graphene, one achieves a value of $D \approx 29 \ev$, which compares well with other estimates of this energy \cite{2008PhRvLBolotin}. 

In the presence of spatially varying density of $\pi$-electrons $\delta n(\vec{x})$, the deformation energy is just $\int d^2\vec{x} D u_{ii}(\vec{x})\delta{n}(\vec{x})$. \footnote{Note that charge neutrality demands $\int d^2 \vec{x} \delta n(x)=0$.}. 

The structure of the electronic membrane is thus determined by the following free energy: 
\begin{eqnarray} \label{Eq:Free_energy}
F[u,h,\delta n] &=&E_{ee}[\delta n]+\frac{1}{2}\int d^2\vec{x}\kappa (\nabla^2 h)^2 + 
\\ \nonumber 
 &+& \frac{1}{2}\int
d^2\vec{x}\left[2\mu u_{ij}^2 +\lambda u_{ii}^2 +2D  u_{ii} \delta n \right]. 
\end{eqnarray}
Where $E_{ee}$ is the electron-electron energy due to the charge density.
% Evidently, the theory is gaussian in the in-plane deformations, thus these degrees of freedom can be integrated out (i.e., solved exactly). This results in a free energy which depends only on the charge density variation and the out-of-plane deformation:
%\begin{eqnarray} \label{Eq:Free_energy_no_u}
%\lefteqn {F_{eff}[h,\delta n] =E_{ee}[\delta n]-\frac{1}{2}\int d^2\vec{x}\frac{D^2}{\mu+\lambda} \delta n^2(x)+}\\ \nonumber &&\frac{1}{2}\int d^2\vec{x}\left\{\kappa (\Delta h)^2+K_0\left[\Delta^{-1}\left(S(\vec{x})-c(\vec{x})\right)\right]^2 \right\}. 
%\end{eqnarray}
%Where $S[h(x)]=\frac{1}{2}(\Delta\delta_{ij}-\partial_i \partial_j)\partial_i h\partial_j h$ is the Gaussian curvature, and $c(\vec{x}) \equiv -\frac{D}{2(\mu+\lambda)}\Delta\delta n(x)$ ($\Delta$ is the Laplacian). The ground state is determined by a minimization of this free energy. 
The free energy evidently couples between between elastic deformations and charge inhomogeneities. %electron-electron interaction energy (first line of Eq.~(\ref{Eq:Free_energy_no_u})), and a part which couples the electronic density and the height fluctuations. 

In order to understand the behavior of this system, let us first concentrate in the electron-electron interaction. 
%The first fact one notices is that the electron-electron interaction includes a part which favors charge fluctuations. This part originates from the deformation energy, as explained mathematically above. It competes with the electrostatic energy, which increases due to charge fluctuations. Such competition leads to phases with non-zero charge fluctuations. i.e., electron-hole puddles will be formed on the lattice. In order to determine whether a charge fluctuation is favored here, one has to know the functional form of the electron-electron interaction. 
Characterizing this interaction is not a trivial task, as it includes a solution of a strongly correlated many body problem. However, one can estimate it by: 
\begin{equation}
E_{ee}=\frac{e^2}{2{\cal E}}\int\int d^2{x}d^2{y} \frac{\delta n(\vec{x}) \delta n (\vec{y})}{|\vec{x}-\vec{y}|}=\frac{2\pi e^2}{{2\cal E}}\int \frac{d^2\vec{q}}{(2\pi)^2}\frac{|\delta n(\vec{q})|^2}{q}, 
\end{equation} 
with $e$ the charge of the electron, the static dielectric constant ${\cal E}$ (originates from screening of the electron-electron interaction), and $\delta n(\vec q)$ a Fourier transform of $\delta n(\vec{x})$. The value of the static dielectric constant is not fully known, since the effective fine structure constant in graphene is of order unity. Recent perturbative analysis of the electron-electron screening by Kotov {\it et al.} \cite{2008PhysRevBKotov} has shown that the static dielectric constant is  ${\cal E} \approx 3-4$. However, they demonstrated that the perturbative analysis receives large corrections at higher orders of perturbation theory, concluding that additional screening is expected in the non-perturbative solution. Here, I will take this value as a starting point for the analysis, and examine the effect of increasing the value of the static dielectric constant on the final conclusions.

The theory is now quadratic in the charge fluctuations, thus can be written as:
\begin{eqnarray} \label{Eq:Free_energy_ee}
F[u,h,\delta n] &=&{\frac{1}{2}\int \frac{d^2\vec{q}}{(2\pi)^2}\left\{\frac{2\pi e^2}{{\cal E}q} \left|\delta n(q)+\frac{D{\cal{E}}q}{2\pi e^2}u_{ii}\right|^2 +\right.}  \nonumber\\ &+&\kappa q^4| h(\vec{q})|^2+\left. 2\mu |u_{ij}|^2 +\lambda(q)|u_{ii}|^2 \right\},
\end{eqnarray}
where $\lambda(q) \equiv \lambda - \frac{D^2 \cal{E}}{2\pi e^2}q$. The theory is gaussian also in the in-plane deformation field $\vec{u}$, thus these degrees of freedom can be integrated out, resulting in an effective free energy that depends only on the out-of-plane deformations:
\begin{eqnarray} \label{Eq:Free_energy_h}
F_{eff}[h] =\frac{1}{2}\int \frac{d^2\vec{q}}{(2\pi)^2}\left\{\kappa q^4|h|^2+K(q)\Phi^2 \right\}. 
\end{eqnarray}
Where $\Phi[h(x)]=\frac{1}{2}(\delta_{ij}-\frac{\partial_i \partial_j}{\nabla^2})\partial_i h\partial_j h$ and $K(q)$ is an effective Young modulus, given by:
\begin{equation} \label{Eq:K}
K(q)=K_0\frac{1-\frac{2\mu+\lambda}{\mu+\lambda}\frac{q}{q_0}}{{1-\frac{q}{q_0}}}
\end{equation} 

Clearly, for long wavelengths (corresponding to $q \rightarrow 0$) $K(q) \rightarrow K_0$, implying that long wavelength sound waves can be used to measure the elastic constants, as they appear in Eq.~(\ref{Eq:Elastic_energy}), neglecting the effects of possible formation of charge inhomogeneities. 

In order to investigate the possibility of ripple excitations, one is interested in the behavior of the electronic crystalline membrane at finite wavelengths. Observing Eq.~(\ref{Eq:K}), it is clear that interesting phenomena occur around the length scale $\xi_0 = \frac{2\pi}{q_0} \equiv \frac{D^2{\cal{E}}}{e^2(2\mu+\lambda)}$. In particular, for $q_0>q>\frac{\mu+\lambda}{2\mu+\lambda}{q_0}$, the effective Young modulus is negative, thus representing a true competition with the bending energy $\kappa (\nabla^2 h)^2$, allowing height fluctuations. In order to quantify this, one has to solve the thermodynamics of the system described by Eq.~(\ref{Eq:Free_energy_h}). 

In general, Dyson equations can be written for the scale evolution of the effective bending rigidity $\kappa_R(q) \equiv (\beta q^4\langle |h_q|^2\rangle)^{-1}$, and Young modulus $K_R(q)$: 
\begin{eqnarray} 
\frac{\kappa_R(q)}{\kappa}& = & 1+\left(\frac{q_T}{q}\right)^2\Sigma(q) 
\\ \label{eq:K_R}\left(\frac{K_R(q)}{K_0}\right)^{-1} &=&  \left(\frac{K(q)}{K_0}\right)^{-1}+\frac{1}{2}\left(\frac{q_T}{q}\right)^2\Psi(q).
\end{eqnarray}
Here, $q_T=\sqrt{\frac{K_0k_BT}{\kappa^2}}$, $\Sigma(q)$ is the sum of all $1PI$ two-point diagrams, and $\Psi(q)$ is the sum of all $1PI$ four-point diagrams. These equations determine the structure of the electronic crystalline membrane. Evidently, two intrinsic length scales exist, i.e., $q_0$, originating in the electronic degrees of freedom, and $q_T$, in which thermal effects become significant. The existence of the scale $q_0$ differentiates electronic crystalline membranes from regular crystalline membranes, whose structure is controlled only by the thermal scale. In the current case, however, there is an interplay between the two length scales. 

As aforementioned, $K(q)\rightarrow K_0$ for $q\ll q_0$. Hence, in this limit, the Dyson equations reduce to the usual case of regular crystalline membranes, extensively studied particularly in the thermodynamic limit $L\rightarrow \infty$ ($q\rightarrow 0$) (see, e.g., \cite{Membranes_Book,2001PhRBowick,1987JPhNelson,1988EPLDavid_1,1988EPLDavid_2, 1988PRLAronovitz, 1992PRLSCSA, 1993SciSchmidt, 1996JPhBowick,2009ArXivGazit}). As a result, electronic crystalline membranes, for which this limit corresponds to the regime $q\ll \{q_0,\, q_T\}$, inherit the same behavior at long wavelengths, i.e., have a stable, asymptotically flat, phase, with the $1PI$ diagrams governed by anomalous exponents, viz. $\Sigma(q)\sim q^{2-\eta}$ and $\Psi(q)\sim q^{2-\eta_u}$, with $\eta\approx0.8$ and $\eta_u=2-2\eta$.

However, the behavior outside this regime, and especially about the electronic scale $q_0$, is governed by the special functional structure of the Young modulus, cf. Eq.~(\ref{Eq:K}). In particular, Eq.~(\ref{eq:K_R}) shows that the formal structure of the effective Young modulus survives the scale evolution, i.e., there always exists a region of negativity for the young modulus, with a zero at $q=\frac{\mu+\lambda}{2\mu+\lambda}q_0$ and a singularity around $q\approx q_0$. As a result, the physics at $q\sim q_0$ will be governed by this length scale, rather than $q_T$. Due to this, as well as the fact that finite $q$ behavior is of interest, we calculate $\Sigma(q)$ in the one-loop approximation ($\cal P$ denotes principal value):
\begin{equation}
\Sigma(q)={\cal P}\int \frac{d^2\vec{k}}{(2\pi)^2} \frac{K(qk)}{K_0} \frac{|\hat{q}\times\hat{k}|^4}{|\hat{q}-\vec{k}|^4}.
\end{equation} 
We search for maximum in the normal-normal correlation function: $G(q) \equiv \langle |\hat{n}_q|^2 \rangle=k_BT/(\kappa_R(q)q^2)=(\kappa/K_0) [(q/q_T)^2+\Sigma(q)]^{-1}$ , as it indicates enhanced correlation, which will manifest itself as ripples on the membrane \cite{2007NatMaFasolino}, if the maximum is pronounced, i.e., if its width is smaller than the characteristic wavelength. The amplitude of the ripples is determined by the maximal value of the correlation. 

\begin{figure}
\resizebox{5.8cm}{!}{\includegraphics[clip=false]{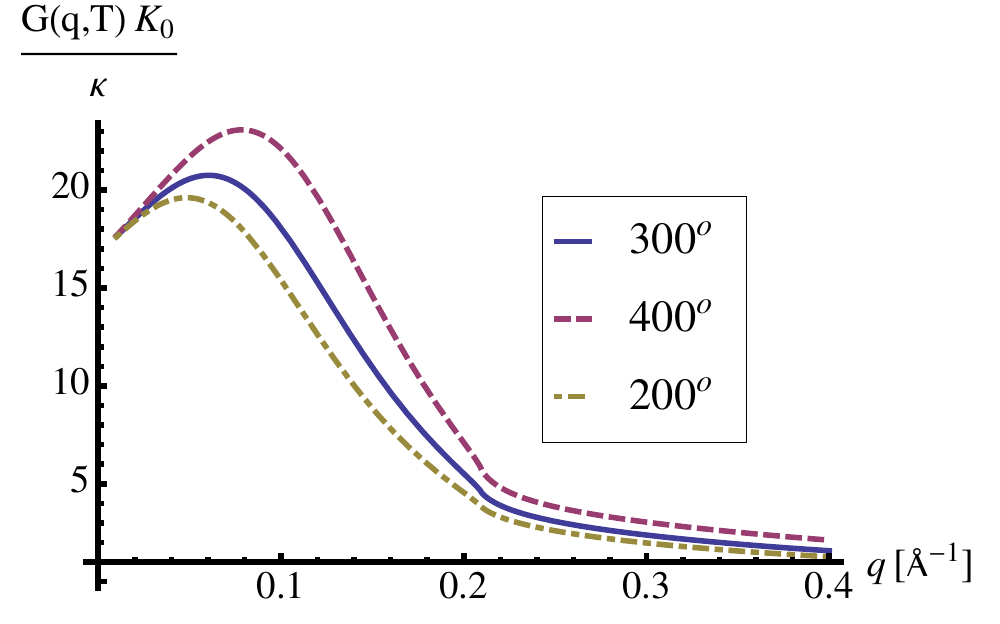}}
\caption{(Color online) The normal-normal correlation function as a function of the wave-number $q$ for different temperatures.}
\label{Fig1}
\end{figure}

The resulting dimensionless normal-normal correlation function $K_0G(q)/\kappa$, as shown in Fig.~\ref{Fig1} for various temperatures, presents a rather pronounced peak. The main temperature dependence is in $q_T$, since simulations suggest rather weak temperature dependence of the elastic constants and the bending rigidity \cite{2009PhysRevLettZakharchenko}. Evidently, the temperature dependence of $G(q)$ vanishes for $q\approx 0.02 \Ang^{-1}$. This signatures the onset of the long-wavelegth regime, where the one-loop approximation given here is expected to fail. Comparing the value of the correlation function at this wavelength to its maximal value is a measure of the amplitude of the ripples. It can be seen that the ripples are more pronounced for higher temperatures, and have shorter wavelength. The width of the correlation maximum weakly depends on the temperature, as it is mainly affected by the electronic length scale $q_0$, rather than the thermal one. The temperature dependence should be further investigated, taking into account changes in other effects, e.g., defect formation and elastic constants. 

\begin{figure}
\resizebox{5.8cm}{!}{\includegraphics[clip=true,viewport=2.8cm 19.5cm 12.5cm 26cm]{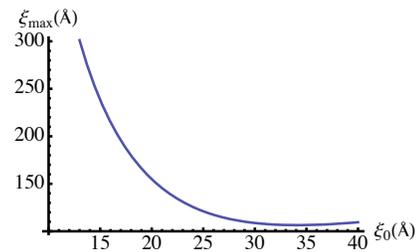}}
\caption{(Color online) Ripples wavelength as a function of the parameter $\xi_0=\frac{D^2{\cal E}}{e^2(2\mu+\lambda)}$.}% whose value holds significant uncertainty, as discussed in the text.}% due to uncertainties in the values of the deformation energy $D$ and the static dielectric constant $\cal E$.}
\label{Fig2}
\end{figure}

$\xi_0=\frac{2\pi}{q_0}$ holds the largest uncertainty in the model, due to the static dielectric constant. Thus, in Fig.~\ref{Fig2} the wavelength maximizing the correlation function is plotted as a function of $\xi_0$ (at room temperature), for a physical range of values of this parameter, chosen around $\xi_0(D=29 \ev,{\cal E}=4) \approx 20\Ang$. As seen in the plot, the maximal correlation occurs for wavelengths in the range $\xi=100-300 \Ang$. This wavelength region reproduces the ripples found in experiments on suspended graphene \cite{2007NatMeyer,2007ApplPhysLettBrar,2009PhysRevLettGeringer}. 

Thus, one expects a stable, asymptotically flat, phase which exhibits charge fluctuations and ripples, with the same correlation length. The character of this correlation can be analyzed through  Eqs.~(\ref{Eq:Free_energy_ee}-\ref{Eq:Free_energy_h}), where the charge fluctuations and in-plane deformations were integrated out. The average values of these integrated out observables are readily recovered, and a relation between the average charge fluctuation and the average lattice deformation is established $\langle \delta n \rangle \propto \langle S[h(x)] \rangle$, where $S[h]=-\nabla^2 \Phi$ is the Gaussian curvature of the surface. This is a key difference between the current work and previous theoretical studies of the correlation between charge puddles and height fluctuations, e.g., Ref.~\cite{2008EPLKim}, that found that charge fluctuations are proportional to the mean curvature, i.e., to $\nabla^2 h$ (indeed, a recent experiment \cite{2009PhysRevBDeshpande} did not find such a correlation).

In suspended graphene, no experiment has studied both the topology and the charge inhomogeneity together. However, the correlation lengths of the two disorder phenomena were measured in different experiments, finding ripples with characteristic size of $100-300 \Ang$ \cite{2007NatMeyer,2007ApplPhysLettBrar,2009PhysRevLettGeringer}, and a correlation length of $300 \Ang$ for charge puddles \cite{2008NatPhMartin}. In the latter experiment, though graphene was suspended on top of SiO$_2$, it was shown that the substrate had no effect on the structure of the charge puddles.
Experiments that probed both phenomena simultaneously were accomplished only for graphene on top of SiO$_2$ substrate \cite{2009PhysRevBDeshpande,2009unpubZhang}, but have shown significant substrate effects. It is important to note that in a setting where the graphene is located very close to the substrate, its structure and its charge fluctuations would be pinned to the surface structure and impurities on the substrate, and not as discussed here. In addition, the current work discusses neutral graphene. Doping is analogue to external stress whose sign corresponds to the majority charge carriers, thus decreasing (increasing) the rippling in the case of electron (hole) doping \cite{2009PhysRevBGazit}. This effect can explain the results of Ref.~\cite{2009unpubZhang}, where the graphene sheet was doped by an external gate voltage. A different external source for ripples in graphene is adsorption of molecules \cite{2009EPLThompson, 2009PhysRevBGazit}. The current work is different since it proposes the electrons as an intrinsic source for ripples at thermodynamic equilibrium.

In conclusion, charge puddles and ripples in graphene are found to be a signature of the fact that graphene is not a regular crystalline membrane, but the herald of a new class of materials -- electronic crystalline membranes, demonstrating strong interplay between the dynamics of the free electrons in the membrane and its mesoscopic structure. Clearly, this implies that had the $\pi$ electrons not been free, the ripples would vanish. Indeed, graphane, an insulating graphene derivative in which each carbon atom is connected to a hydrogen atom, was found to exhibit reduced corrugations \cite{2009ScienceElias}.

This paper offers a theoretical approach to characterize the two main intrinsic disorder phenomena in graphene, i.e., charge inhomogeneity and structural deformations. As graphene is promising material for technological use, understanding disorder phenomena and correlations among them is essential for a successful design and quality control of future applications.

%This leads, in addition, to a simple formula for the electron-phonon coupling in superconductors, hinting to competition with the appearance of charge density fluctuations correlated to lattice deformations. 

%In general, electronic crystalline membranes are predicted to be in a unique state, characterized as a glassy phase of charge inhomogeneities and corrugations. Charge inhomogeneity coupled to in-plane deformation is found possible even when out-of-plane deformations are suppressed, thus providing a possible explanation to the correlation found between bond stretching anomalies and charge inhomogenieties observed in some layered materials that show high temperature superconductivity (see, e.g., \cite{2006NatureReznik}). Whether this enhanced electron-phonon interaction can lead to better understanding of high temperature superconductivity is remained to be explored.

%As graphene is a promising material for technological use, the theoretical procedure presented here provides an important tool for characterizing its intrinsic structure -- a key element in the design and quality control of any future application. 

%{\it Acknowledgments -- }
%%%%%%%%%%%%%%%%%%%%%%%%%%%
I thank K. Novoselov, L. Radzihovsky, Dam T. Son, A. Fasolino and G. Bertsch for helpful discussions. This work was supported, in part, by DOE under grant no. DE-FG02-00ER41132. 
%%%%%%%%%%%%%%%%%%%%%%%%%%%

%\bibliography{graphene-bibliography-3-shortened}{}
\end{document}